# A Comprehensive Regime Diagram of Dynamical Modes of Triple Flickering Buoyant Diffusion Flames: Experimental and Model Investigations

Hanxu Wang[1,2], Tao Yang[3,4], Yicheng Chi[5], Zhenyu Zhang[1,2,*], Peng Zhang[3,4,*]

1. School of Mechanical Engineering, Beijing Institute of Technology, Beijing, China
2. National Key Laboratory of Science and Technology on Multi-Perch Vehicle Driving Systems, Beijing Institute of Technology, Beijing 100081, P. R. China
3. Department of Mechanical Engineering, City University of Hong Kong, Kowloon Tong, Kowloon, Hong Kong
4. Shenzhen Research Institute, City University of Hong Kong, Shenzhen, 518057, P. R. China
5. School of Automotive and Transportation Engineering, Shenzhen Polytechnic University, Shenzhen 518055, PR China

**Abstract**

The triple-flame system serves as the fundamental unit for understanding multi-flame interactions, revealing critical coupling mechanisms that scale to complex burner arrays. In this study, we investigated triple flame oscillators, consisting of three flickering laminar buoyant diffusion flames arranged in an isosceles triangular configuration, to construct a comparative regime diagram of dynamical modes. To overcome the limited experimental observability caused by the discretization of geometric parameters, we enabled continuous motion of the vertex flame at a controlled speed $V$, while independently varying the base length $L$ and the fuel flow rate $Q$. We conducted a systematic investigation of the triple flame coupling behaviors by varying the triangle size, fuel flow rate, and vertex flame movement velocity. Based on the experimental observations, a comprehensive regime diagram was established to classify the dynamical modes of triple flickering buoyant diffusion flames. Notably, three previously unreported dynamical modes were identified for the first time. To interpret these complex flame interactions, a Stuart-Landau oscillator model with time-delay coupling was employed, which successfully reproduces the experimentally observed dynamical modes. Experimentally observed dynamical modes reveal a bifurcation diagram for the coupled triple Stuart-Landau system, elucidating the transitions between different synchronization modes.

---
* Corresponding author
  E-mail address: zhenyu.zhang@bit.edu.cn (Z. Zhang); penzhang@cityu.edu.hk (P. Zhang)

# 1. Introduction

Combustion instability remains a critical challenge in modern combustion systems [1, 2]. Although the mechanisms of single-burner instability have been extensively studied, the complex dynamics of multi-burner configurations, such as annular arrays in gas turbines [3] and clustered nozzle arrangements in rocket engines [4], are far less understood [5]. Fundamental flame systems, characterized by well-defined coupling dynamics, provide simplified and insightful models for exploring the behavior of larger flame systems [6]. Regime diagrams play a crucial role in bridging fundamental understanding and practical engineering in flow and flame fields [7-11], as they can present a clear overview of the relation between each flame pattern and the operational parameters.

As a fundamental component of many domestic and industrial systems, diffusion flames often exhibit self-excited flickering instabilities, driven by the periodic formation and detachment of toroidal vortices around the top of the flame [12, 13]. The flicker of flame is a fundamental manifestation of buoyancy-driven flow dynamics and is relevant to many combustion fields, including wake flames [14], soot formation [15], and flames under external forcing [16, 17]. Owing to their reproducible and well-characterized oscillatory behavior, flickering flames have emerged as an ideal model system for investigating unsteady combustion dynamics [18-23].

Extensive studies [23-28] have established that the fundamental buoyancy-driven vortex mechanism governs the behavior of coupled flames, providing a rigorous physical basis for investigating synchronization and mode transitions. As a pioneering work, Kitahata et al. [29] developed a theoretical framework for both individual and coupled flame oscillators, successfully reproducing characteristic candle flame dynamics. Bunkwang et al. [30] employed combined experimental and numerical approaches to systematically compare single- and dual-flame systems, demonstrating their underlying physical similarity while revealing distinct frequency jump phenomena during axisymmetric mode transitions. Yang et al. [31] numerically performed detailed analyses of vortex dynamics in coupled buoyant diffusion flames, offering deeper insight into the fundamental mechanisms of synchronization.

The introduction of a third flame in triangular configurations substantially broadens the range of observable coupling phenomena. In triple flickering diffusion flame systems, the emergent dynamics are strongly influenced by inter-flame spacing.

Okamoto et al. [33] first reported four distinct flickering modes (i.e., in-phase, oscillation death, rotational, and partial in-phase) by varying the geometric arrangement of triangular candle flame arrays. More recently, Yang et al. [28] reproduced these four modes numerically and provided physical interpretations based on the vortex interaction processes, including vorticity reconnection and vortex-induced flows. Chi et al. [34] conducted systematic experiments on triple flickering flames in fixed isosceles configurations, identifying seven stable dynamical modes, and subsequently proposed a regime diagram for the triplet-flame system [35]. These studies underscore the importance of comprehensive parametric investigations, particularly including geometric configurations and flame variations, to fully characterize complex coupling dynamics.

However, previous studies on triple flames in triangular configurations have typically explored only a limited set of geometric parameters, resulting in an incomplete characterization of the full range of dynamical modes. Notably, the preliminary regime diagram proposed by Chi et al. [35] exhibits certain limitations in distinguishing between mode boundaries, with some regions ambiguously labeled as a mixed mode—a consequence of the geometric constraints imposed by the chessboard-type burner platform. In addition, few studies [23, 36] have attempted to systematically explore dynamical models capable of replicating experimental findings. To date, a comprehensive experimental and modeling study has yet to be conducted to investigate the dynamical modes and mode transitions of triple flames across a continuous range of configurations, hindering a comprehensive understanding of their rich and intricate coupling behaviors.

To address this gap, we propose a novel experimental approach in which the vertex flame in triangular arrangements is displaced at a controlled, low velocity, allowing continuous variation of geometric parameters. This enabled a systematic and comprehensive investigation of synchronization dynamics in triple flickering flames. Furthermore, a time-delay Stuart Landau oscillator model is employed to reproduce the experimentally observed dynamical modes. The resulting comprehensive regime diagram not only obtained previously reported behaviors but also reveals new dynamical modes, providing a physical framework that could inform real-time geometric control strategies in multi-injector combustion systems.

The remainder of this work is organized as follows: Section 2 describes the experimental setup for the triple flickering flames with a horizontally moving vertex flame, along with the mode recognition methods. Section 3 discusses the effect of horizontal movement on a single flickering flame in experiments and numerical simulations. Then examines the influence of vertex flame displacement on the flickering frequency and coupling boundaries, which confirms the complete regime diagram. After introducing the oscillator model and demonstrating its ability to replicate the experimental observations, three new dynamical modes are identified. In the process of simulation, a bifurcation diagram of triple Stuart-Landau oscillators was obtained. Finally, Section 4 summarizes the main findings, discusses their significance, and outlines potential applications.

## 2. Experimental and Theoretical Methodology

### 2.1 Experimental Setup

Figure 1 shows a schematic of the experimental setup and a top view of the geometry of the flame arrangement. The triple flickering buoyant diffusion flames are arranged in an isosceles triangle configuration, which encompasses both the straight-line and equilateral-triangle flame arrangements reported in previous studies [24, 37]. Three identical Bunsen-type burners (each square tube with a side length $d = 10$ mm and a height $h = 120$ mm) were fixed at the same height on an adjustable stand. The nozzle outlet was slightly pinched to minimize the effect of tube wall thickness on the flame base. These burners were fueled by gaseous methane (99.9% purity) at a flow rate of $Q = 0.3, 0.4, 0.5,$ and $0.6$ L/min, covering the range typically for flickering buoyant diffusion flames in previous studies [22, 34, 38]. Flow rate $Q$ was precisely regulated by the mass flow controller (Alicat Scientific, MC-Series:5SLPM-D/5 M) to ensure all three flames were identical. Flame behavior was recorded using a high-speed camera (FASTCAM Nova S12, Photron) operating at 125 frames per second. To eliminate depth-of-field effects in measuring the flame properties (e.g., flame size and brightness) from snapshots, the camera was positioned 1 m away from the flame array, a distance much greater than the characteristic flame height (~10 cm).

As shown in Fig. 1(b), the three identical flames (denoted by Flame L, C, and R) were arranged in an isosceles triangle with variable base length $L$ and height $H$. Previous studies [24, 34] have demonstrated that variations in these geometric parameters ($L$ and $H$) can induce distinct dynamical modes in the triple flame system. In the present study, the base length was set to $L = 45, 50, 60,$ and $70$ mm, respectively, as significant coupling between two adjacent flames occurs within approximately $7d$ [34]. To enable continuous $H$, the vertex flame (Flame C) was fixed on the slider of an electronically controlled linear stage, with displacement speeds $V =$ 2.5, 5.0, and 7.5 mm/s over a maximum distance of $H_{max}=$ 160 mm. In this configuration, the triangle height increases with time $t$ as $H(t) = V \cdot t$, allowing the flame geometry to evolve continuously during each run. This design permits systematic mapping of mode boundaries across a wide range of triangular configurations, overcoming the discrete parameter limitations of previous studies [24, 34].

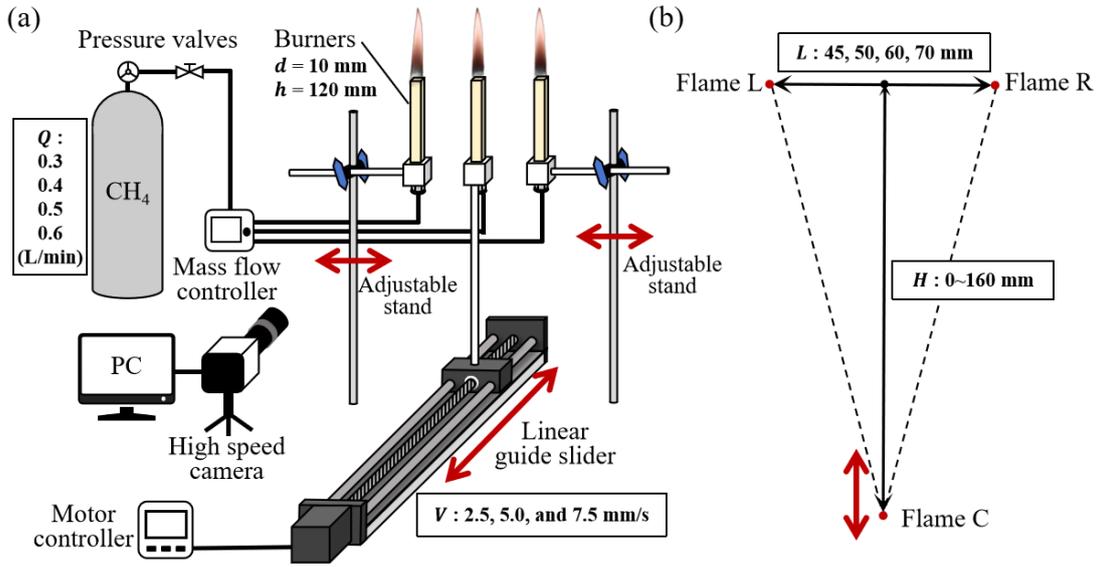

Fig. 1. Schematic of the triple-flame system with a horizontally moving vertex flame: (a) experimental setup and (b) isosceles triangular arrangement of the three flames. Methane-fueled diffusion flames are generated on three identical square burners. The two base flames (Flame L and Flame R) are fixed at a separation distance $L$, while the vertex flame (Flame C) is mounted on a motor-controlled linear slider and moves at a constant velocity $V$, resulting in a continuously varying triangle height $H$.

A parametric study was conducted by varying $Q$, $L$, and $V$ to investigate dynamical modes of triple flickering buoyant diffusion flames. To facilitate the following discussion, we used non-dimensional parameters of Reynolds number $Re = Ud/\nu_F$ and Froude number $Fr = U^2/gd$, where $U = Q/d^2$ is the inlet velocity from each burner, $\nu_F = 1.71 \times 10^{-5}$ m²/s is the kinematic viscosity of gaseous methane at 20°C and 1 atm, and $g = 9.8\ m/s^2$ is the acceleration of gravity. The parameter ranges explored in this work are summarized in Table 1. Under these conditions, each Bunsen-type burner fueled by gaseous methane produces a buoyant diffusion flame, whose flickering (or puffing) results from buoyance-induced instability at small Froude numbers ($Fr \ll 1$) [39-42]. The dynamical modes of the triple flames were identified through analysis of the flame behaviors in both physical and phase spaces, following approaches established in previous studies [34, 35].

Table 1. Ranges of experimental parameters for triple flickering flames.

| Parameter | $Q$ (L/min) | $U$ (mm/s) | $L$ (mm) | $V$ (mm/s) | $Re$ | $Fr$ |
|---|---|---|---|---|---|---|
| Ranges | 0.3~0.6 | 50~100 | 45~70 | 2.5~7.5 | 29~59 | $2.55\times10^{-2}$ ~$1.02\times10^{-1}$ |

## 2.2 Phenomenal Reproduction and Dynamical Modes

Identification of dynamical modes in coupled flame systems requires selecting time-dependent variables that characterize both individual flame behavior and inter-flame interactions. Many previous studies [26, 34, 43-45] have shown that the flickering dynamics of buoyant diffusion flames can be depicted using either local parameters (e.g., pressure, temperature, luminosity) or global parameters (e.g., morphology, brightness). Same as previous studies [24, 34, 38, 46], we employ flame brightness extracted from front-view snapshots, omitting three-dimensional collective effects; this simplification is justified as no significant differences were observed between single and triple flickering flames. Therefore, a time-dependent variable, the brightness value of each flame $B_i(t)$, $i = L, C, R$, is extracted from flame videos for our subsequent analysis.

We have successfully reproduced six distinct dynamical modes of triple flames arranged in an isosceles triangle configuration, as recently reported in previous work [34]. Representative flame snapshots and the corresponding normalized brightness signals for these typical modes (Modes I–VI) are presented in Fig. 2(a) and 2(b), respectively. The modes are defined as follows:

I. In-phase mode: All three flames oscillate periodically in synchrony, including simultaneous pinch-off events. The normalized brightness signals $\tilde{B}_L$, $\tilde{B}_C$, and $\tilde{B}_R$ oscillate in phase, maintaining an approximate zero-phase difference ($\Delta\phi = 0$).

II. Flickering death mode: Oscillatory motion is strongly suppressed, with minimal amplitude and no observable pinch-off. The normalized brightness signals $\tilde{B}_L$, $\tilde{B}_C$, and $\tilde{B}_R$ remain near zero.

III. Partially flickering death mode: Flame L and Flame R exhibit anti-phase flickering ($\Delta\phi_{LR} = \pi$) with periodic pinch-off, while Flame C remains nearly steady ($\tilde{B}_C(t) \approx 0$) and shows no pinch-off.

IV. Partially in-phase mode: Flame L and Flame R flicker in-phase ($\Delta\phi_{LR} = 0$), whereas Flame C flickers out-of-phase ($\Delta\phi = \pi$) with respect to them. Notably, all three flames exhibit pinch-off behavior.

V. Rotation mode: The three flames flicker sequentially with a stable right→center →left (R-C-L) phase progression ($\Delta\phi \approx 2\pi/3$ between successive flames), forming a persistent rotational mode.

VI. Decoupled mode: The three flames exhibit asynchronous flickering with time-varying phase differences ($\Delta\phi$), irregular pinch-off timing, and non-uniform oscillation amplitudes among $\tilde{B}_L$, $\tilde{B}_C$, and $\tilde{B}_R$.

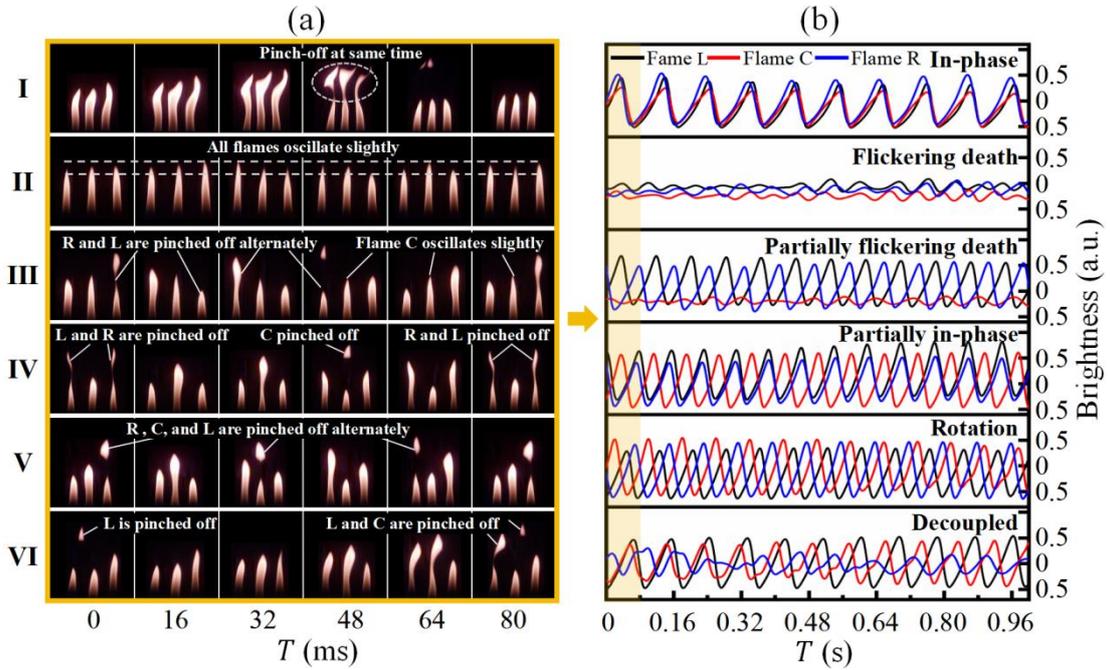

Fig. 2. Six dynamical modes of the triple flame system: (a) representative flame snapshots recorded by the high-speed camera and (b) corresponding normalized brightness signals $\tilde{B}_L$, $\tilde{B}_C$, and $\tilde{B}_R$. Mode I: in-phase mode ($Q = 0.5\ L/min$, $L = 50\ mm$, $H = 0.3{\sim}5.3\ mm$); Mode II: flickering death mode ($Q = 0.4\ L/min$, $L = 70\ mm$, $H = 13.1{\sim}18.1\ mm$); Mode III: partially flickering death mode ($Q = 0.4\ L/min$, $L = 70\ mm$, $H = 72.0{\sim}77.0\ mm$); Mode IV: partially in-phase mode ($Q = 0.4\ L/min$, $L = 70\ mm$, $H = 45.4{\sim}50.4\ mm$); Mode V: rotation mode ($Q = 0.4\ L/min$, $L = 50\ mm$, $H = 41.5{\sim}46.5\ mm$); Mode VI: decoupled mode ($Q = 0.4\ L/min$, $L = 60\ mm$, $H = 146.9{\sim}149.4\ m$).

The formation of these dynamical modes can be interpreted through vortex-dynamics mechanisms [28]. For example, in in-phase mode, vortex reconnection around the three flames generates a 'trefoil' vortex structure. The periodic shedding of this trefoil vortex induces simultaneous necking and pinch-off of the three flames. In flickering death mode, the trefoil vortex detaches at the downstream of flames, suppressing and preventing pinch-off. In contrast, rotation mode is characterized by the alternate shedding of toroidal vortices without significant vorticity reconnection, in this case, vortex-induced flows dominate the flame-flame interactions and drive the observed sequential flickering behaviors.

## 2.3 Instantaneous Phase and Frequency of Flickering Flames

In this study, $B(t)$ was calculated in the same way with our previous work [35]. In brief, each color snapshot (each pixel containing RGB values) was first converted into grayscale (pixel intensity range of 0~255). The three flames were segmented individually, and the grayscale intensity within each flame region was integrated to yield $B_i(t)$ (for $i = L, C, R$). To facilitate subsequent analysis, the brightness was normalized brightness

$$\tilde{B}_i(t) = [B_i(t) - \bar{B}_i(t)]/\bar{B}_i(t) \tag{1}$$

where $\bar{B}_i(t)$ is the temporal mean of $B_i(t)$. This normalization centers the oscillations around zero while preserving relative amplitude and phase information. The resulting time series $\tilde{B}_L$, $\tilde{B}_C$, and $\tilde{B}_R$ are then analyzed over at least ten consecutive flickering cycles to identify dynamical modes. Additional details of the brightness extraction process are provided in the Supplementary Material.

The various dynamical modes of triple-coupled flames arise from the complex interaction of the vortex rings surrounding the flames [28]. Quantitative characterization of these modes requires determination of the instantaneous phase and frequency of flickering. The analytic signal framework provides an effective means to extract these instantaneous quantities from a time series, capturing the signal's local characteristics at each point in time. In particular, the Hilbert transform enables a direct computation of the instantaneous phase and its temporal derivative, thereby yielding the instantaneous frequency [47].

In this study, the instantaneous phase and frequency of $\tilde{B}_i(t)$ were determined using the Hilbert Transform, which has been shown to be particularly effective for analyzing the dynamic characteristics of flame brightness [35, 48]. The Hilbert Transform $H[\tilde{B}_i(t)]$ of a real-valued $\tilde{B}_i(t)$ is defined as

$$H[\tilde{B}_i(t)] = -\frac{1}{\pi}\text{p.v.}\sum_{-\infty}^{\infty}\frac{\tilde{B}_i(\tau)}{t-\tau}d\tau \tag{2}$$

where 'p.v.' denotes the Cauchy principal value, used to handle the singularity at $t = \tau$. Then we can convert the real-valued signal into a complex-valued analytic signal as

$$s_i(t) = \tilde{B}_i(t) + jH[\tilde{B}_i(t)]. \tag{3}$$

from which the instantaneous phase $\phi_i(t)$, the phase difference $\Delta\phi_{ij}(t)$ and the instantaneous frequency $f_i(t)$ are calculated as

$$\phi_i(t) = \arctan\left(\frac{H[\tilde{B}_i(t)]}{\tilde{B}_i(t)}\right) \tag{4}$$

$$\Delta\phi_{ij}(t) = \text{unwrap}[\phi_i(t) - \phi_j(t)], \quad i,j \in \{L, C, R\} \tag{5}$$

$$f_i(t) = \frac{1}{2\pi}\frac{d\phi_i(t)}{dt} \tag{6}$$

where 'unwrap' means that the angle is shifted by adding multiples of $\pm 2\pi$ until the phase jump is less than $\pi$. A detailed description of the Hilbert transform implementation is provided in the Supplementary Material.

## 2.4 Stuart-Landau Model of Flame Oscillator

To interpret the experimental observations, we employ a dynamical model that approximates the complex flame–flow system, described in full by partial differential equations (PDEs), with a reduced-order set of ordinary differential equations (ODEs) that retains the essential dynamical features. Previous experiments[49-51] demonstrated that flame flickering exhibits a nearly sinusoidal periodic oscillation, which can be modeled as a self-sustained nonlinear oscillator [35, 37]. Oscillator models are widely used to study the behavior of both artificial and natural systems, including vortex shedding in the wake of a cylinder [52].

Recently, Yang et al. [23] qualitatively modeled the dynamical modes of octuple flickering flames in circular arrays using Stuart-Landau (S-L) oscillators, which describe the behavior of a nonlinear oscillating system near the Hopf bifurcation. The S-L oscillator has proven to be a simple and effective approach for modeling flickering flames [35, 46], and is given by

$$\frac{dZ(t)}{dt} = (a + i\omega - |Z(t)|^2)Z(t) \tag{7}$$

where $Z(t) = r(t)e^{i\phi(t)} = Re(Z) + i\text{Im}(Z)$ is a complex variable quantity representing the oscillation with amplitude $r(t)$ and phase $\phi(t)$, $a$ is the bifurcation parameter (oscillations occur only for $a > 0$), and $\omega$ is the natural frequency.

The interaction of flames is modeled by the coupling term of S-L oscillators. For the triple-flame system, we use three coupled S-L oscillators, $Z_i(t)$ with $i = L, C, R$, including a time-delay term characterized by coupling strength $K$ and time delay $\tau$ [23, 48]. The coupling strength $K$ is associated with flame size (larger flames correspond to a large $K$), while the time delay $\tau$ reflects the flame separation distance[48]. For the isosceles triangle configuration, we define $\tau_1$ for the base length and $\tau_2$ for the waist length. Assuming identical flames ($K$ identical for all pairs), the governing equations are

$$\frac{dZ_L}{dt} = (a_L + i\omega_L - |Z_L|^2)Z_L + K(Z_C(t - \tau_2) - Z_L) + K(Z_R(t - \tau_1) - Z_L) \quad (8)$$

$$\frac{dZ_C}{dt} = (a_C + i\omega_C - |Z_C|^2)Z_C + K(Z_L(t - \tau_2) - Z_C) + K(Z_R(t - \tau_2) - Z_C) \quad (9)$$

$$\frac{dZ_R}{dt} = (a_R + i\omega_R - |Z_R|^2)Z_R + K(Z_L(t - \tau_1) - Z_R) + K(Z_C(t - \tau_2) - Z_R) \quad (10)$$

with $a_L = a_C = a_R = 1$ and $\omega_L = \omega_C = \omega_R = 10$ to represent a limit-cycle state without loss of generality.

A parametric study of the model is performed by varying $K$, $\tau_1$, and $\tau_2$ to correspond to changes in flame size and geometry in the experiments. To account for the environmental disturbance introduced by horizontal flame motion, we also include a Gaussian white-noise term of intensity of $\epsilon$ on the right-hand side of Eqs. (8)-(10), yielding a noisy S-L oscillator system. The corresponding MATLAB code is provided in the Supplementary Material.

## 3. Results and Discussions
### 3.1 Single Flickering Flames with Horizontally Moving

The flickering frequency of a single buoyant diffusion flame $f_s$ follows the well-known scaling law of $f_s \sim (g/d)^{1/2}$, which arises from the periodic shedding of the buoyancy-induced vortices [43]. To examine the influence of the horizontal motion, we carried out a series of investigations under different fuel flow rates $Q$ and moving speed $V$ of the single flame in Fig. 3. The flickering behaviors of a single flame ($Q = 0.4$ L/min) under both quiescent conditions ($V = 0$ mm/s) and moving conditions ($V = 2.5, 5.0,$ and $7.5$ mm/s) were illustrated in Fig. 3(a). For comparison, a benchmark case is included from our previous work [34], which examined a flickering flame from a circular Bunsen-type burner in a quiescent condition.

In all cases ($V = 0 \sim 7.5$ mm/s), each flickering cycle involves gradual flame elongation, necking at mid-height, and eventually pinch-off into two separate parts, with a typical variation of flame height at $\sim 4d$. Qualitative comparison demonstrates that both burner geometry and translational speed have a negligible effect on buoyancy-induced flickering behavior, with no significant variation observed in flame morphology.

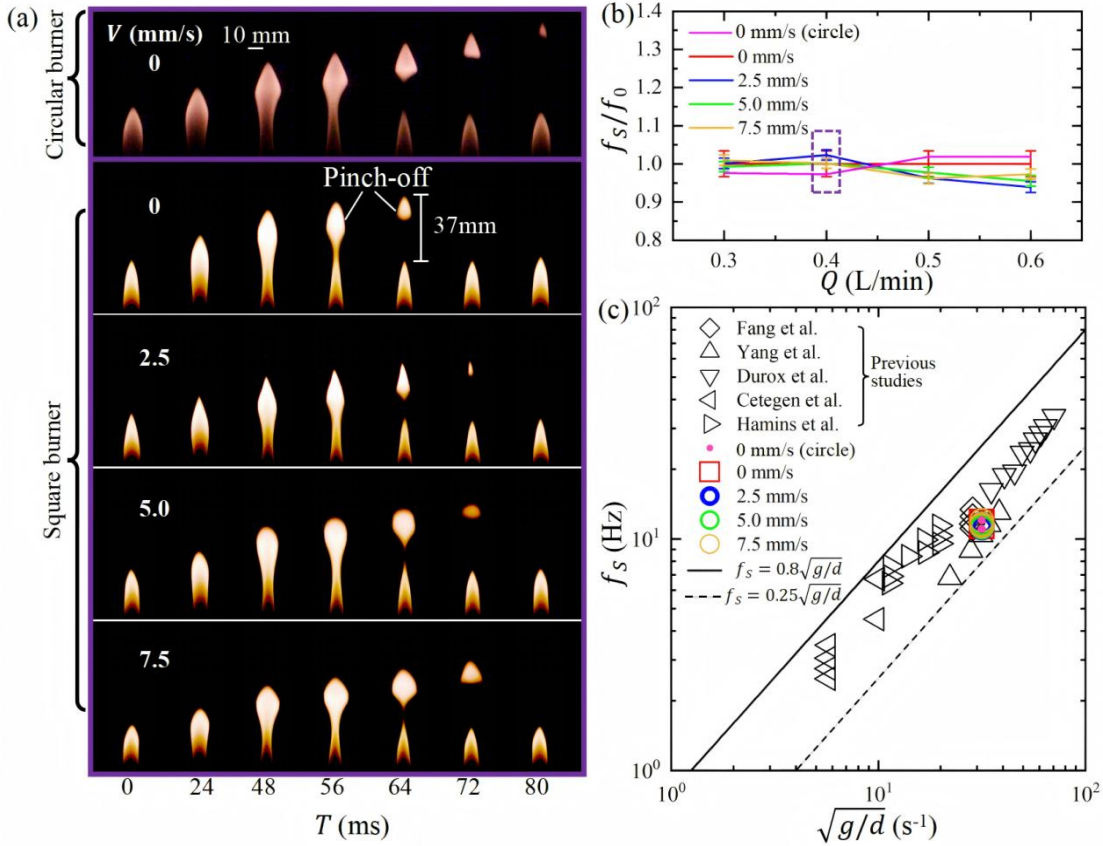

Fig. 3. The flickering characteristics of a single laminar buoyant diffusion flame under different geometric and translational conditions: (a) Sequential snapshots of a single flame at $Q = 0.4$ L/min for different burner geometries and $V$. The experiment with the circular burner was done by Chi et al. [34]. (b) The normalized flickering frequency $f_s/f_0$ for single flame under varying $Q$, $V$, and nozzle shape. (c) The scaling law of the flickering frequency of the single flame.

To facilitate a quantitative comparison, we further examined the normalized flickering frequency $f_s/f_0$ for single diffusion flames under varying $Q$ and $V$, where $f_s$ is the measured frequency at different fuel flow rates and $f_0$ denotes the flickering frequency under the quiescent conditions, as shown in Fig. 3(b). The results show that $f_s/f_0$ remains close to unity, with variations within $\pm 0.05$, indicating that horizontal translation has only a minimal influence on the flickering frequency. In Fig. 3(c), we examine the present experiments with previous studies [28, 43, 53-55] for the scaling law. All data collapse onto the scaling relation $f_s = C(g/d)^{1/2}$, with a prefactor $C$ in the range of 0.25~0.80, demonstrating good agreement. This finding indicates that the flame flickering observed in this study aligns with the well-established mechanism of buoyancy-driven vortex shedding [13].

In this study, we noticed that the characteristic velocity of flame height variation is approximately 400 mm/s (estimated by $4d \cdot f_0$), which is more than an order of magnitude greater than the imposed translational velocity (< 10 mm/s). This order-of-magnitude difference implies that the flickering of horizontally translating flames can be characterized as a quasi-steady oscillatory process. Additionally, we observed that horizontal motion induces only minor variations in $f_s$. Therefore, in the modeling framework, the influence of $V$ is treated as an external stochastic disturbance, represented by a Gaussian noise term with intensity $\epsilon$ in the S-L oscillator equations, which is investigated in Fig. 4.

As shown in Fig. 4(a-b), the sawtooth waveform of the brightness signal $\tilde{B}$ of flickering flames can be qualitatively simulated by a sinusoidal wave of a S-L oscillator. The results indicate that both translational velocity $V$ and noise strength $\epsilon$ have negligible influences on the shape and amplitude of periodic oscillations. However, changes in $V$ may induce a phase shift in an individual flame, which could be attributed to the influence of horizontal flow on the vortex shedding dynamics in the vertical direction. A detailed discussion of these effects will be presented in Section 3.2. Figure 4(c) shows $f_s/f_0$ of the S-L oscillators remains essentially unchanged for $\epsilon = 0\sim10$, which is consistent with the experimental results in Fig. 3(b).

These comparisons justify that the single horizontally translating flames could be qualitatively modelled by S-L oscillators with an additive noise term for investigating the dynamical modes of multi-flame systems [23]. The sinusoidal waveform corresponds to the projection of a limit cycle, which evolves from the initial state of $(0,0)$ in the complex plane, as shown in Fig. 4(d-e). For small noise levels $\epsilon \leq 1$, the trajectories converge smoothly into a closed orbit, whereas for larger noise $\epsilon > 1$, the limit-cycle trajectories exhibit obvious fluctuations, but ultimately also concentrate to a closed orbit.

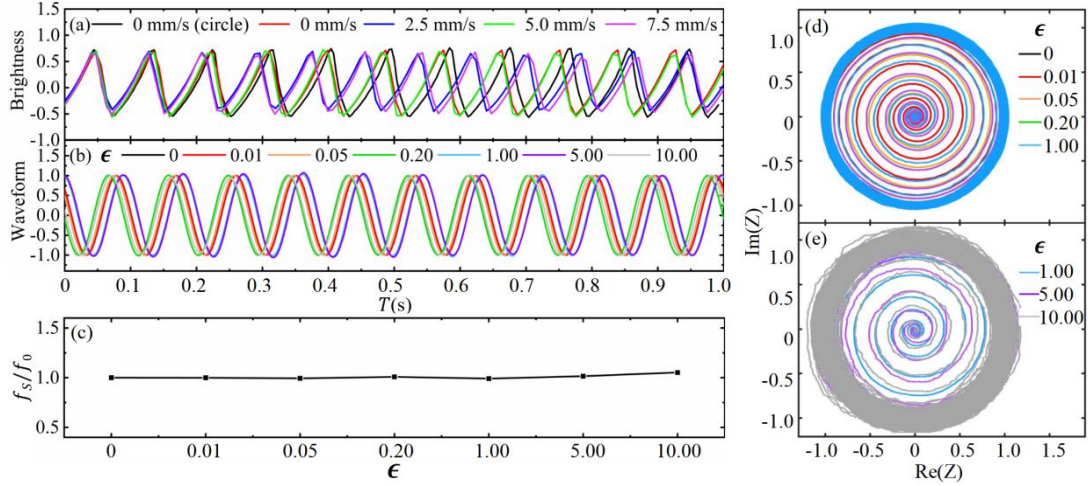

Fig. 4. Effect of horizontal translation and noise strength on single flame oscillations. (a) The normalized brightness signals of a single flame at $Q = 0.4$ L/min for translational velocities $V = 0$, 2.5, 5.0, and 7.5 mm/s (including a circular-burner as reference). (b) The periodic waveforms of a single S-L oscillator under different noise strength $\epsilon$. (c) The normalized oscillation frequency ($f_s/f_0$) of the S-L oscillator under different $\epsilon$. (d-e) Development of oscillator limit cycles in the complex plane for small and large values of $\epsilon$.

## 3.2 Comprehensive Regime Diagram of Triple Flickering Flames

The instantaneous phase and frequency are used to study the dynamic features of triple flickering flames, as illustrated in Section 2.3. When the vertex flame is fully decoupled from the two base flames, its flickering frequency $f_c$ approaches the frequency $f_s$ of single flame, while the phase difference between the vertex flame and the two base flames becomes indeterminate. To analyze the effect of $V$ on synchronization and mode transition in triple flame systems, we present the normalized frequency ratio of $f_c/f_s$ in Fig. 5(a-c) and the phase difference in Fig. 5(d-f). Same with the previous work [35], the dimensionless parameter space of $Gr^{0.5}\Delta L(1 + \alpha Fr)^{-1}$ and $\Delta H(t)/\Delta L$ is employed for regime diagram of triple flickering flames, where $Gr = gd^3/v_F^2$ is the Grashof number, the correction coefficient $\alpha = 1.2$, $\Delta H = H/d$ is the dimensionless instantaneous height, and $\Delta L = L/d$ is the dimensionless base scale.

In Fig. 5(a-c), the lower-left region of the regime diagrams, of which boundary is the black contour line of $f_c/f_s = 1.0$, indicates strong interaction among the triple flames, as their distances are quite close and the coupling frequency is slightly lower than $f_s$. When $H$ increases at a relatively large, $f_c$ becomes higher than $f_s$, which corresponds the upper-left region of the regime diagrams. As the vertex flame moves away, the coupling strength gradually decreases, and the system transitions toward weaker synchronization. The red dashed lines denote the transition boundary between strong and weak coupling regimes, where the frequency has an abrupt descent. These frequency trends, which initially rise and subsequently fall, resemble those observed in dual flame systems. Notably, these transition boundaries remain essentially unchanged across different $V$ from 2.5, 5.0, and 7.5 mm/s, indicating that horizontal motion within this range has an insignificant influence on the regime transitions of triple flame systems. However, it was observed that the phase difference in triple flames seems to have a dependence on $V$, especially the frequent changes within weakly coupled regimes, as shown in Fig. 5(e-f). This dependence is likely attributed to the effect of $V$ on slight phase shifts of flickering flames, as illustrated in Fig. 4(a).

In this section, we focus on the regime diagram of triple flames at a translational speed of $V = 5.0$ mm/s, and compare it with the diagram reported by the previous work [35] in Fig. 6. Additional regime diagrams corresponding to other speeds are provided in the Supporting Material. As the previous diagram is consisted of sparse points, we plot a simple schematic to facilitate comparison, using the color-coded mode regions, as shown in Fig. 6(a). The extended parameter ranges explored in the present experiments fill gaps in previously unknown regions. The distributions of various dynamical modes largely are consistent within the black dashed boxes in Fig. 6(a-b), indicating that synchronization and mode transitions in an isosceles triangular array with a translating vertex flame exhibit strong similarity to those in the static configuration. Moreover, the present study complements key jigsaw pieces in the existing regime diagram through more detailed data, thereby offering a more comprehensive understanding of the dynamic modes in triple flame systems.

Regarding the mode distribution in the regime diagram in Fig. 6(b), our results indicate that strongly coupled modes (Modes I–V) are predominantly situated to the left of the coupling transition lines. As the vertex flame moves farther away and their coupling weakens, decoupled mode (Mode VI) emerges. Importantly, three new dynamical states of asymmetric partially flickering death mode (Mode III-2), death decoupling mode (Mode VI-2), and asymmetric partially in-phase mode (Mode IV-2) were identified in our experimental configuration. These newly observed modes can persist for over ten consecutive flickering cycles and recur across varying parameter ranges. Their detailed characteristics will be described in the following sections.

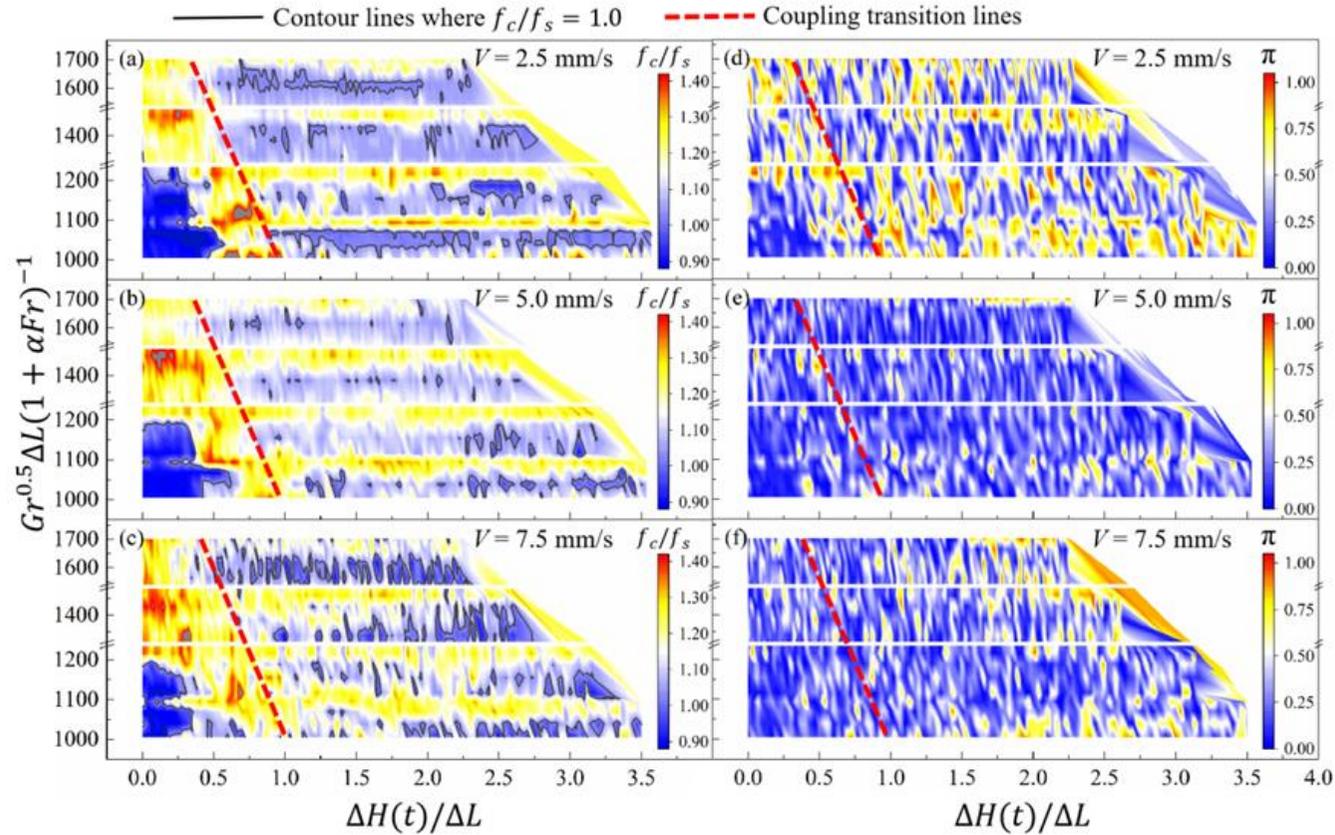

Fig. 5. The coupling characteristics of triple flickering flames in an isosceles triangle configuration, represented in the dimensionless parameters space of $Gr^{0.5}\Delta L(1 + \alpha Fr)^{-1}$ and $\Delta H/\Delta L$. (a-c) The normalized frequency ratio $f_c/f_s$ of vertex flame under the moving speeds $V = 2.5$, 5.0, and 7.0 mm/s, respectively. The red dashed lines separate strong and wake coupling regions, which remain essentially unchanged with increasing $V$. (d-f) The average phase difference between vertex flame and the two base flames under the moving speeds of $V = 2.5$, 5.0, and 7.0 mm/s, respectively.

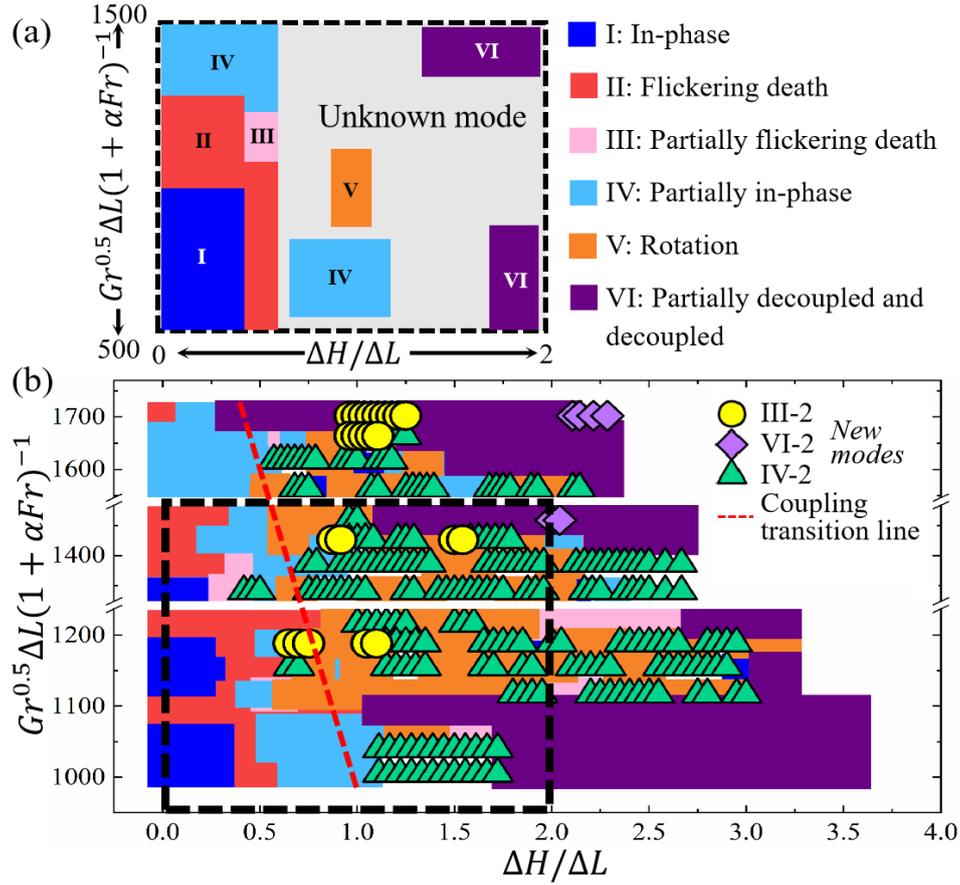

Fig. 6. The regime diagrams of triple flame systems in the dimensionless parameter space of $Gr^{0.5}\Delta L(1+\alpha Fr)^{-1}$ and $\Delta H/\Delta L$. (a) Dynamical modes of Mode I-VI observed in previous studies [35, 37] and the grey area is unidentified. (b) Dynamical modes identified in the present study for $V = 5.0$ mm/s. The dashed black boxes indicate the same parameter ranges.

### 3.3 Phase Portraits of Dynamical Modes in Experiment and Model

Corresponding characteristic behaviors of triple flames in physical space in Section 2.2, we present three-dimensional phase portraits of $\tilde{B}_L$, $\tilde{B}_C$, and $\tilde{B}_R$ of Mode I-VI in experimental and numerical results along with their two-dimensional projections ($\tilde{B}_L - \tilde{B}_C$, $\tilde{B}_L - \tilde{B}_R$, and $\tilde{B}_C - \tilde{B}_R$) in Fig. 7. The temporal evolution of flame dynamics is indicated by the color bar. In general, the S-L model successfully captures the general phase-space trajectories of Modes I–VI observed in our experiments, qualitatively reproducing the distinct dynamical behaviors of the triple flame system. However, quantitative differences between the experimental and modeled patterns are observed, which we attribute to the presence of oscillations with "N"-shaped and sinusoidal wave characteristics, as shown in Fig. 4. The S-L model is capable of qualitatively simulating the sawtooth signal of single flickering flames as a sinusoidal wave.

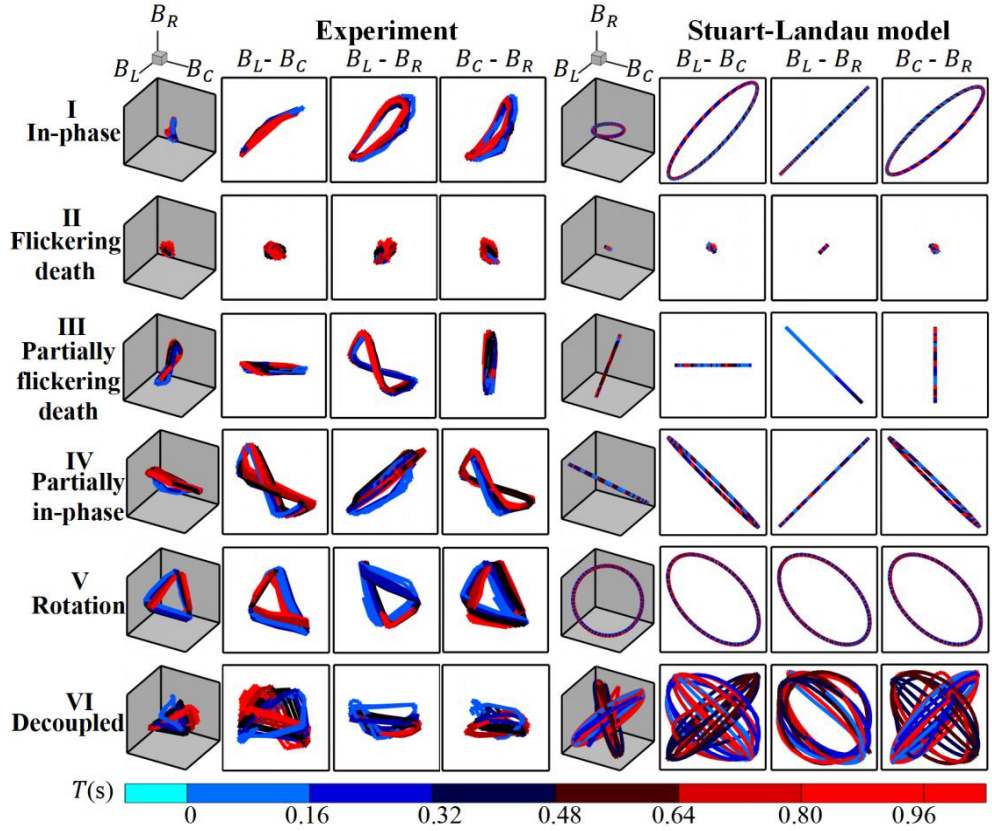

Fig. 7. Phase trajectories of six typical dynamical modes in the triple flame system from the present experiment (left) and Stuart-Landau oscillator model (right). The detailed parameters are listed: Mode I: in-phase mode ($Q = 0.5\ L/min$, $L = 50\ mm$, $H = 0.3 \sim 5.3\ mm$; $K = -3.99$, $\tau_1 = 4.60$, $\tau_2 = 2.60$); Mode II: flickering death mode ($Q = 0.4\ L/min$, $L = 70\ mm$, $H = 13.1 \sim 18.1\ mm$; $K = 5.00$, $\tau_1 = 3.37$, $\tau_2 = 2.28$); Mode III: partially flickering death mode ($Q = 0.4\ L/min$, $L = 70\ mm$, $H = 72.0 \sim 77.0\ mm$; $K = 0.45$, $\tau_1 = 0.30$, $\tau_2 = 0.20$); Mode IV: partially in-phase mode ($Q = 0.4\ L/min$, $L = 70\ mm$, $H = 45.4 \sim 50.4\ mm$; $K = 2.88$, $\tau_1 = 4.50$, $\tau_2 = 4.10$); Mode V: rotation mode ($Q = 0.4\ L/min$, $L = 50\ mm$, $H = 41.5 \sim 46.5\ mm$; $K = 0.25$, $\tau_1 = 2.20$, $\tau_2 = 2.20$); Mode VI: decoupled mode ($Q = 0.4\ L/min$, $L = 60\ mm$, $H = 146.9 \sim 149.4\ mm$; $K = -2.37$, $\tau_1 = 0.60$, $\tau_2 = 1.73$).

These phase-space trajectories, corresponding to these six dynamical modes, are compared and discussed as follows:

I. In-phase mode: In the experiments, the three flames flicker synchronously with negligible phase differences. Their trajectory forms a slender ellipsoid oriented along the diagonal (1,1,1) vector in phase space, with all three 2D projections appearing as ellipses aligned with (1,1). In the S-L model, all flame oscillators behave without any phase difference, producing consistent topological structures in phase space.

II. Flickering death mode: The experiment and model both show that all projections collapse to a point approximately, indicating a negligible variation over time.

III. Partially flickering death mode: Both the experiment and the Stuart-Landau model exhibit two horizontal 2D projections and a negative diagonal 2D projection, as the vertex flame ceases to flicker and the base flames are in anti-phase. It is noteworthy that the Stuart-Landau model successfully reproduces a phase trajectory spread along the direction (1, -1) for $B_L - B_R$, resembling the butterfly-shaped pattern observed experimentally.

IV. Partially in-phase mode: The experiment shows an ellipse-shaped 2D projection aligned with (1,1) in $B_L - B_R$ and two butterfly-shaped 2D projections aligned with (1, -1) in $B_L - B_R$ and $B_L - B_R$, indicating that the two base flames flicker in phase with each other, but alternately with the vertex flame. The model reproduces this dynamical behavior in phase space with geometrically regular patterns. In Model IV, the $D_2$ symmetry of the isosceles triangle configuration constrains that the two vertex-base flame pairs necessarily exhibit identical dynamical behavior.

V. Rotation mode: Experimental phase trajectory has three 2D triangle projections, reflecting a phase difference of $2\pi/3$ between each pair of flames. In the model, the sinusoidal oscillations of flame oscillators yield three identical ellipses, corresponding to the same phase difference.

VI. Decoupled mode: In both experiment and model, no interactions occur among those flames, resulting in completely uncorrelated phase relationships. Due to the ergodicity of the decoupled system, the two-dimensional projections fails to form any repeatable patterns.

As a result, three-dimensional phase portraits can effectively capture the essence of dynamical modes through their distinct topological geometries, thereby facilitating the identification of characteristic behaviors in new modes.

## 3.4 New Death Modes

As reported in Section 3.2, three new dynamical modes of triple flickering buoyant diffusion flames were identified in our comprehensive regime diagram. In this section, we provide a detailed analysis of Mode III-2 and Mode VI-2, in which one or two flames exhibits slight oscillations without the flame pinch-off (e.g., flickering death [34,

48]). Their dynamical features in physical and phase spaces are recognized.

As shown in Fig. 8, Mode III-2 is characterized by alternating the pinch-off between Flame L and Flame C with a phase difference of $\pi$, while Flame R exhibits slight oscillations in a flickering death state. Interestingly, its phase portraits resemble those of Mode III, but the symmetry constraint of the isosceles triangular is broken in Fig. 8(a). Therefore, this dynamical behavior is denoted as the asymmetric partially flickering death mode. The normalized brightness signals in Fig. 8(b) further highlight the asymmetric role of the base flames, consistent with the snapshots in Fig. 8(c). The distribution of Mode III-2 in the regime diagram is shown in Fig. 8(d), where the three circles represent different $V$ and the black dashed line indicates the transition between strong and weak coupling regimes. Mode III-2 primarily occurs near the transition where coupling strength weakens, leading to a breakdown of $D_2$ symmetry. Particularly, sparse geometric variations in Chi et al. [35] likely led to the oversight of this mode.

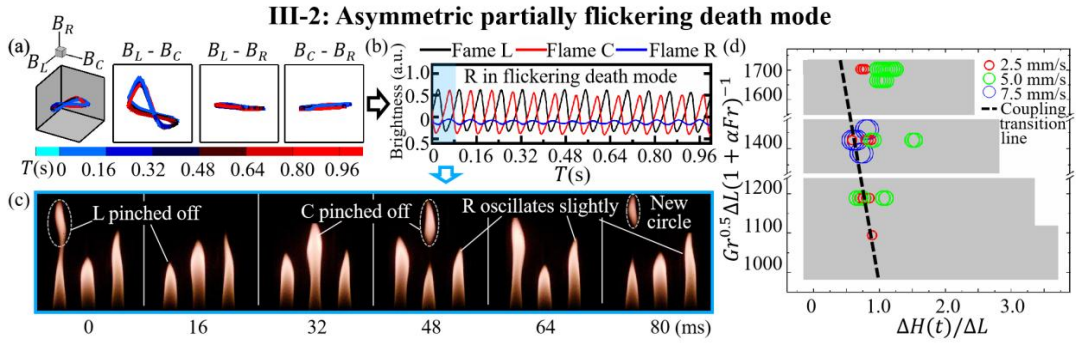

Fig. 8 Mode III-2: asymmetric partially flickering death mode. (a) Three-dimensional phase portraits of $\widetilde{B}_L$, $\widetilde{B}_C$, $\widetilde{B}_R$, along with their two-dimensional projections. (b) Normalized brightness signals ($\widetilde{B}_L$, $\widetilde{B}_C$, and $\widetilde{B}_R$) in about 1s. (c) Flame snapshots of sequential oscillations within one period. (d) Distribution of Mode III-2 under different $V$ in the regime diagram.

To further investigate the repeatability of Mode III-2, we fixed the vertex flame at the observed locations of the triple flames with horizontally moving in Fig. 8 (d). As shown in Fig. 9 (a), this mode was reproduced near the transition line even in the absence of translational speed. An example shows that the three-dimensional phase portraits in Fig. 9(b) have similar topological structures to those in Fig. 8(a). The results strongly confirm that our proposed configuration with a moving vertex flame provides a reliable platform for the systematic study of triple-flame systems. Moreover, it was observed in Fig. 9(a) that Mode III-2 occurs more frequently under translational motion, which could be beneficial to trigger this mode.

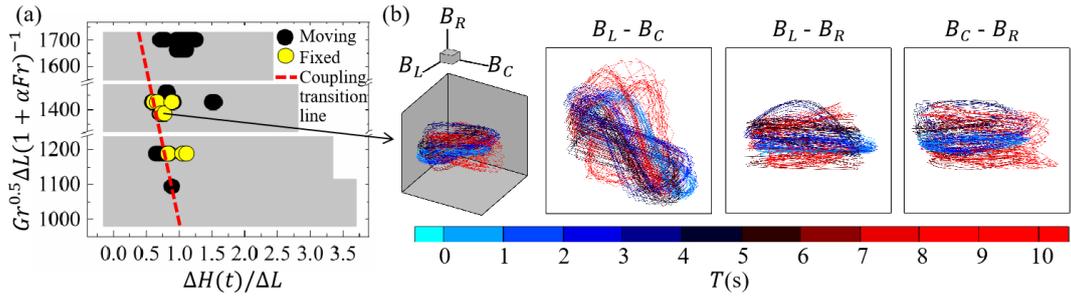

Fig. 9 Phenomenal reproduction of Mode III-2 in fixed configuration of triple flames and their moving counterparts. (a) Mode distribution in the regime diagram. (b) Three-dimensional phase portraits of $\tilde{B}_L$, $\tilde{B}_C$, and $\tilde{B}_R$, along with their two-dimensional projections ($Q = 0.5\ L/min$, $L = 60\ mm$, $H = 4.3\ mm$).

For the death decoupling mode (Model VI-2), phase-space trajectories, time-varying brightness signals, and physical snapshots in Fig. 10(a-c) collectively demonstrate the two base flames remain in a flickering death state whereas the vertex flame flickers independently. In the regime diagram of Fig. 10(d), this mode only occurs when the vertex flame moves sufficiently far from the two base flames. The triple-flame system partially decouples into two distinct subsystems: the base ones, which remain coupled in a flickering death state, and the vertex one, which becomes a single flickering flame.

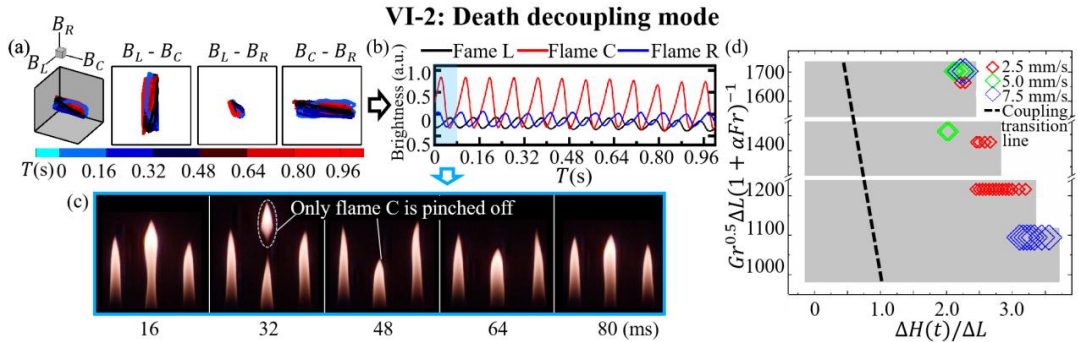

Fig. 10. Mode VI-2: death decoupling mode. (a) Three-dimensional phase portraits of $\tilde{B}_L$, $\tilde{B}_C$, and $\tilde{B}_R$, along with their two-dimensional projections. (b) Normalized brightness signals ($\tilde{B}_L$, $\tilde{B}_C$, and $\tilde{B}_R$) in about 1s. (c) Flame snapshots of sequential oscillations within one period. (d) Distribution of Mode VI-2 in the regime diagram.

### 3.5 Asymmetric Partially In-phase Mode by Experiment and Model

In this section, we provide a detailed analysis of Mode IV-2, in which the vertex flame flickers in phase with one of the base flames in Fig. 11, constituting an asymmetric partially in-phase mode. Interestingly, its phase portraits resemble those of Mode IV, but the symmetry constraint of the isosceles triangular is broken in Fig. 11(a).

The time-varying brightness signals in Fig. 11(b) indicate that Flame C and Flame R oscillate in phase with each other but maintain a phase difference of π relative to Flame L. Representative physical snapshots of this mode are provided in Fig. 11(c). Moreover, Mode IV-2 were observed in a relatively broad region, spanning the strong and weak coupling regimes, in Fig. 11(d).

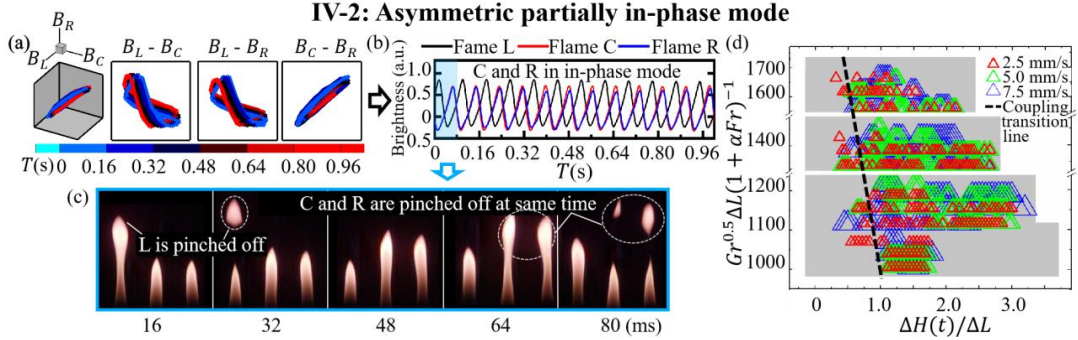

Fig. 11. Mode IV-2: asymmetric partially in-phase mode. (a) Three-dimensional phase portraits of $\widetilde{B}_L$, $\widetilde{B}_C$, and $\widetilde{B}_R$, along with their two-dimensional projections. (b) Normalized brightness signals ($\widetilde{B}_L$, $\widetilde{B}_C$, and $\widetilde{B}_R$) in about 1s. (c) Flame snapshots of sequential oscillations within one period. (d) Distribution of Mode IV-2 in the regime diagram.

In addition, this mode can also be reproduced by the S-L model in Fig. 12(a), where the waveform and phase difference confirm that the vertex flame and one base flame oscillate in phase ($\varphi_{CR} \approx 0$), while both remain out of phase with the other base flame ($\varphi_{LC} \approx \varphi_{LR} \approx \pi$). The S-L model successfully mimics this mode within the parameter ranges of $-0.5 < K < 0.5$ and $0 < \tau_1, \tau_2 < 5$ in Fig. 12(b-e).

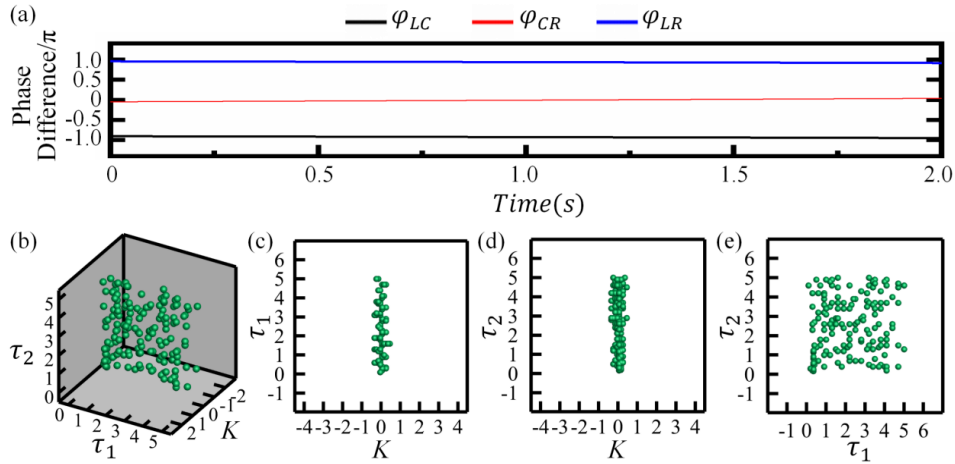

Fig. 12. Mode IV-2: asymmetric partially in-phase mode reproduced by the Stuart-Landau model. (a) Phase difference of three S-L oscillators ($K = 0.25$, $\tau_1 = 2.20$, $\tau_2 = 3.40$). (b) All model parameters of $K$, $\tau_1$, and $\tau_2$ for Mode IV-2 in the ranges of $-0.5 < K < 0.5$ and $0 < \tau_1, \tau_2 < 5$.

It is worth noting that Mode III-2 and Mode VI-2 were not reproduced by the present S-L model. Their absence may be attributed to the inherent limitations of the S-L oscillator model, which is just a simplified representation of flame oscillator (i.e., a toy-model approach for the flickering flame [23, 35, 46, 48]). Therefore, a direct correspondence between experimental variables (e.g., $Q$, $L$, $V$) and model parameters (e.g., $K$, $\tau$, $\epsilon$) has not yet been established in this study and deserves future works.

### 3.6 Bifurcation Diagram of Triple Stuart-Landau Oscillators

To establish the correspondence between experiments and the S-L oscillator model, we performed simulations of the triple oscillator system across a range of coupling parameters $K$, $\tau_1$, and $\tau_2$. Figs. 13(a)-(d) present bifurcation maps obtained for coupling strength values $-2 \leq K \leq 2$, with fixed $\tau_1 = 0.05, 0.20, 0.35$, and $0.50$, while $\tau_2$ was varied from 0.5 to 3.5 times of $\tau_1$. Fig. 13(e) extends the parameter space, exploring a broader domain of $-5 \leq K \leq 5$ and $\tau_2/\tau_1$ up to 8. The $x$-axis was adjusted according to the density of the points because the selected points of the three parameters in the simulation are all discrete and do not cover the entire range.

Within this parameter space, the model successfully reproduced all six previously reported dynamical modes (I-VI), as well as the newly identified asymmetric partially in-phase mode (IV-2). However, the asymmetric partially flickering death mode (III-2) and the death decoupling mode (VI-2), both observed experimentally, did not emerge in the model. This discrepancy highlights limitations of the present oscillator framework, which cannot yet capture all experimentally observed asymmetries and stochastic effects.

Two main factors likely contribute to these differences. First, the S-L oscillator is a reduced-order representation that omits certain vertex interaction mechanisms, such as asymmetric vortex reconnection, which are critical for III-2 and VI-2. Second, only discrete sets of parameters were explored in this study. A more continuous and systematic parameter sweep may reveal additional dynamical regimes. Future refinements of the oscillator model, potentially incorporating higher-order coupling terms of noise-driven bifurcations, will be necessary to fully reproduce the experimentally observed dynamics.

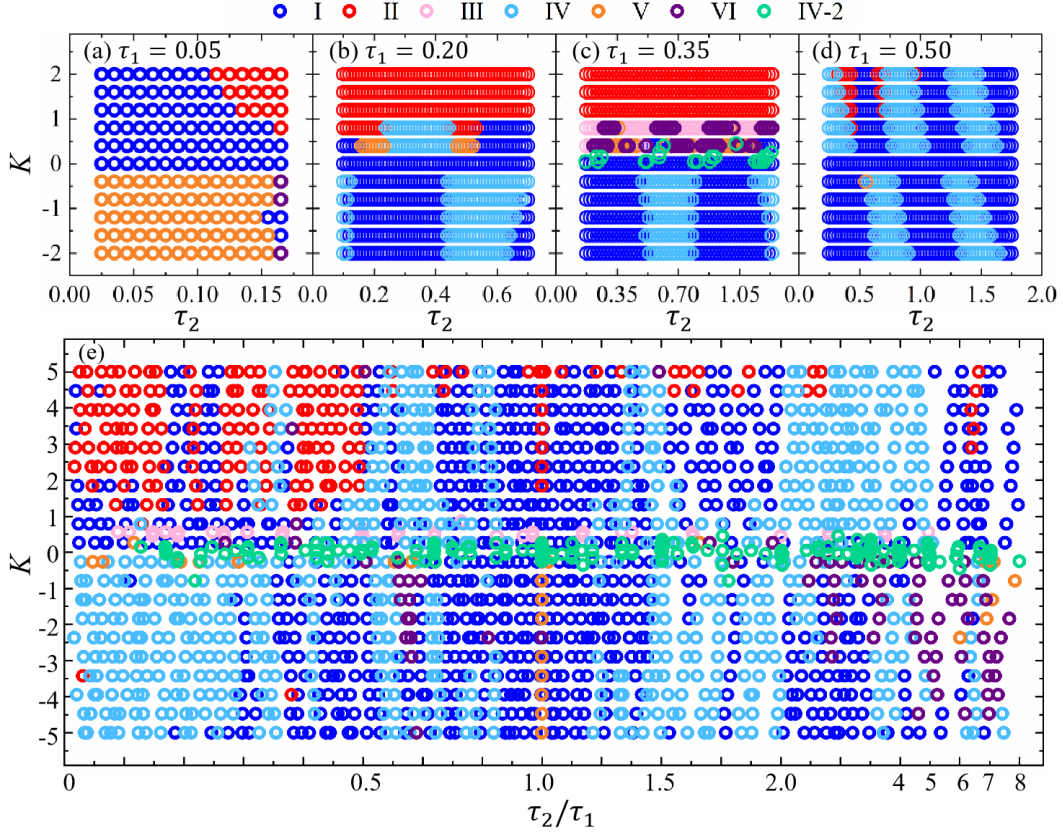

Fig. 13. Bifurcation diagram of triple Stuart-Landau oscillator systems. (a)-(d) Mode distributions in the parameter space of $K$ and $\tau_2$ at a fixed $\tau_1 =0.05, 0.20, 0.35$, and $0.50$, respectively. (e) Comprehensive bifurcation map in the parameter space of $\tau_2/\tau_1$ and $K$.

Although there is no strict one-to-one correspondence between the model oscillator parameters ($K$, $\tau_1$, and $\tau_2$) and the physical parameters in the experiments, several qualitative comparisons can still be made. In Fig. 13(a), Modes V and VI appear when $K$ is negative, while in Figs. 12(b-d), the distributions of Modes I and IV are less concentrated compared to the experimental regime diagram. Notably, in Fig. 13(c), Modes III and IV-2, which are relatively scattered in the distribution of experiment results, only appear when $K < 1$. As shown in Fig. 13(e), the overall bifurcation diagram of the triple Stuart-Landau system does not reproduce the detailed regime structure of the experimental diagram in Fig. 6. Furthermore, $\tau_2/\tau_1 < 0.5$ and $K < 0$, which have no corresponding physical meaning in the experiment.

Nevertheless, meaningful physical correlations can still be identified. Since the time delay $\tau$ is inherently linked to characteristic flame distances, certain parameter ratios capture experimental observations. For example, Mode V occurs predominantly near $\tau_2/\tau_1 = 1.0$, consistent with the experimental finding that the rotation mode is most likely to appear in equilateral triangular configurations. Similarly, Mode VI appears at $\tau_2/\tau_1 > 4.0$, reflecting the experimental observation that decoupling arises when the vertex flame (Flame C) is sufficiently far from the two base flames. These consistencies suggest that, despite its limitations, the Stuart-Landau oscillator retains predictive value for mode classification. In the future, we will explore more continuous parameter sweeps and refine the model formulation to better reproduce experimentally observed asymmetries and stochastic behaviors.

## 4. Concluding Remarks

This study systematically investigated the dynamical behaviors of isosceles triangular triple flickering laminar buoyant diffusion flames using a novel experimental configuration with a continuously movable vertex flame. By enabling precise control of the fuel flow rate $Q$, the base length $L$, and the vertex flame displacement velocity $V$, we were able to overcome the limitations of previous discrete configurations and extend the parameter space of triple flame dynamics. Flame interactions were captured with high-speed snapshots, and their synchronization characteristics were interpreted using a time-delay coupled Stuart-Landau (S-L) oscillator model.

The results demonstrate that horizontal motion of the vertex flame at moving speeds $V = 2.5 - 7.5$ mm/s exerts a negligible influence on the intrinsic flickering frequency of single flames or on the collective coupling frequency of triple flames. A comprehensive regime diagram was established, filling previously uncharted parameter regions and expanding the coverage of dynamical modes. In addition to reproducing six previously reported modes, three new dynamical modes were discovered: Mode III-2 (asymmetric partially flickering death mode), Mode VI-2 (death decoupling mode), and Mode IV-2 (asymmetric partially in-phase mode). The physical mechanism underlying these modes was clarified. Mode III-2 is a transition weak coupling mode, where the vertex flame alternatively pinches off with one base flame while suppressing the toroidal vortex of the other. The movement of the vertex flame can facilitate this mode compared to a fixed system. Mode VI-2 is a partially decoupled mode, in which the vertex flame oscillates independently while base flames remain in flickering death due to mutual suppression. Mode IV-2 is a strong coupling mode, originating from the rotation mode, where the vertex flame becomes phase-locked with one base flame and remains out of phase with the other. Although detailed mechanisms of formation of those modes still remain unclear, the present experimental study produced regime diagrams of existence of different flame patterns.

The S-L oscillator model successfully reproduced six typical modes and Mode IV-2, and a bifurcation diagram of the oscillator network was also established. Although discrepancies remain, particularly in reproducing weak and decoupled modes, the model computation highlights both the potential and the limitations of reduced-order oscillator models in capturing multi-flame dynamics. In the future, we will focus on refining the S-L framework, exploring more continuous parameter combinations, and incorporating additional stochastic or nonlinear interaction terms to achieve closer correspondence with experimental observations.

**Declaration of Competing Interest**
The authors declare that they have no known competing financial interests or personal relationships that could have appeared to influence the work reported in this paper.

**Acknowledgements**
This work is supported by the National Natural Science Foundation of China (No. 52176134) and partially by the APRC CityU New Research Initiatives/Infrastructure Support from Central of City University of Hong Kong (No. 9610601).

**Supplementary Materials**
Supplementary material associated with this article can be found, including the post-processes, noisy S-L oscillator model, and regime diagrams.